\documentclass[twocolumn,epsf,epsfig,showpacs,prd]{revtex4}
\usepackage{graphicx}
\usepackage{dcolumn}
\usepackage{bm}

\usepackage[dvips]{color}
\definecolor{Black}{named}{Black}
\definecolor{Blue}{named}{Blue}
\definecolor{Red}{named}{Red}

\begin{document}

\title{Lorentz Violation and Ultrahigh-Energy Photons}
\author{Matteo Galaverni$^{a,b,c}$, G{\"u}nter Sigl$^{d}$}

%\email{galaverni@iasfbo.inaf.it}
\affiliation{$^a$Dipartimento di Fisica, Universit\`a di Ferrara,
via Saragat 1, I-44100 Ferrara, Italy}
\affiliation{$^b$INAF-IASF Bologna, 
via Gobetti 101, I-40129 Bologna, 
Italy}
\affiliation{$^c$INFN, Sezione di Bologna,
via Irnerio 46, I-40126 Bologna, Italy}
\affiliation{$^d$II.~Institut f\"ur Theoretische Physik, Universit\"at Hamburg,
Luruper Chaussee 149, 22761 Hamburg, Germany}

\begin{abstract}
The propagation of photons, electrons and positrons at ultra-high energies
above $\sim10^{19}\,$eV can be changed considerably if the dispersion relations
of these particles are modified by terms suppressed by powers of the Planck scale.
We recently pointed out that the current non-observation of photons in the
ultra-high energy cosmic ray flux at such energies can put strong constraints
on such modified dispersion relations. In the present work we generalize these
constraints to all three Lorentz invariance breaking parameters that can occur
in the dispersion relations for photons, electrons and positrons at first and second
order suppression with the Planck scale. We also show how the excluded regions in these three-dimensional
parameter ranges would be extended if ultra-high energy photons were detected in the future.
\end{abstract}


\maketitle

\section{Introduction}

Many modern extensions of the Standard Model of particle physics,
including string theory and various other approaches aiming at
unification of quantum mechanics with general relativity, suggest
that the Lorentz symmetry may be broken or modified at energy and
length scales approaching the Planck scale. While such effects necessarily
have to be tiny at energies up to the electroweak scale in order to
satisfy laboratory constraints~\cite{Kostelecky:2008ts}, they can be magnified
at higher energies as they can occur in astrophysics. The observation of standard radiation
processes as they are predicted in the absence of Lorentz invariance
violation then often allows to derive strong constraints on Lorentz
invariance violating (LIV) effects~\cite{Jacobson:2005bg,Bietenholz:2008ni}.

LIV effects could, for example, change
the propagation and thus spectra and composition of the highest
energy particles observed in Nature \cite{Coleman:1998ti,Aloisio:2000cm}. In particular,
photons are produced as secondaries of cosmic rays but are quickly reabsorbed
in the cosmic microwave background (CMB) if Lorentz symmetry holds. If
photon interactions with the microwave background were inhibited by
new physics near the Planck scale $M_{\rm Pl}$, a significant fraction of ultra-high
energy (UHE) cosmic rays above $10^{19}$ eV would be photons, in contradiction
to current experimental upper limits~\cite{Risse:2007sd}. Based on this argument,
in Ref.~\cite{Galaverni:2007tq} we put strong constraints on the coefficients of
LIV terms of the form $(E/M_{\rm Pl})^n$, $n=1,2$,
in the dispersion relation for photons of energy $E$, assuming for simplicity,
that the corresponding parameters for electrons and
positrons are significantly smaller. These latter terms can, however,
play an equally important role because they can influence the kinematics
of pair production, $\gamma\,\gamma_{\rm b}\rightarrow e^-\,e^+$, the most important
process for absorption of high energy photons in the background of low energy
photons $\gamma_{\rm b}$, such as the CMB. In the present work we, therefore, extend
those studies to all three LIV terms that are allowed at first and second order suppression
with the Planck scale in the dispersion relations of photons, electrons and positrons from general
principles of effective field theory, without making any assumptions on the relative size of these
parameters. We obtain three-dimensional exclusion plots for the coefficients of these
terms, following from the non-observation of photons above $\simeq10^{19}\,$eV. 
We also show how the excluded regions in these three-dimensional
parameter ranges would be extended if ultra-high energy photons were detected
in the future. Throughout this paper we assume that primary cosmic rays are dominated by protons.

For the coefficients of LIV terms linearly suppressed with the Planck scale
values larger than $\sim 10^{-5}$ for electrons and positrons are currently 
ruled out by the observation of synchrotron radiation from the Crab Nebula \cite{Maccione:2007yc}.
We find that the observation of photons above $\simeq10^{19}\,$eV would rule out
that any one of the three LIV coefficients for electrons, positrons and photons has
absolute value $\gtrsim 10^{-14}$. This is in agreement with Ref.~\cite{Maccione:2008iw}
which considers a two dimensional subset of the general three-dimensional parameter range.

In contrast, we will find that for LIV terms quadratically suppressed with the Planck scale,
arbitrarily large values of one of the LIV terms for electrons and positrons can not be ruled
out by UHE photon observations if the coefficients of the two other
LIV terms have absolute value $\lesssim10^{-6}$.

The remainder of this paper is structured as follows: In Sect.~II we introduce
the modified dispersion relations, in Sect.~III we determine the thresholds
for pair production and photon decay, and in Sect.~IV we derive the resulting
constraints on the LIV parameters. Finally, we summarize and conclude in
Sect.~V. We use natural units, $c=1$, throughout.

\section{Modified dispersion relations}
We indicate the 4-momenta for ultrahigh-energy photons with $(\omega,{\bf k})$, 
for low-energy background photons with $(\omega_{\rm b},{\bf k_{\rm b}})$,
for electrons with $(E_{\rm e},{\bf p_{\rm e}})$,
and for positrons with $(E_{\rm p},{\bf p_{\rm p}})$.
  
The following modifications to the Lorentz invariant dispersion relations are considered:
\begin{eqnarray}
\omega^2_\pm&=&k^2+\xi^{\pm}_{n}k^2\left(\frac{k}{M_{\rm pl}}\right)^n\,,\label{eq:disp_ph}\\
\omega_{\rm b}^2&=&k_{\rm b}^2\,,\\
E_{\rm e,\,\pm}^2&=&p_{\rm e}^2+m_{\rm e}^2+\eta^{\rm e,\,\pm}_{n} p_{\rm e}^2\left(\frac{p_{\rm e}}{M_{\rm pl}}\right)^n\,,\label{eq:disp_e}
\end{eqnarray}
where $n\geq 1$, $M_{\rm pl}\simeq 10^{19}$ GeV is the Planck mass, $m_{\rm e}\simeq0.5$ MeV is the electron mass, 
and the $+\left(-\right)$ sign in Eq.~(\ref{eq:disp_ph}) for photons indicates right (left) polarization, 
while in Eq.~(\ref{eq:disp_e}) for electrons denotes positive (negative) helicity.

The LIV parameters $\xi^{\pm}_{n}$ for the two photon polarizations are not independent;
they are related by effective field theory considerations \cite{Myers:2003fd,Mattingly:2008pw}
as $\xi_n^+=(-1)^n \xi_n^-$ for $n=1,\,2$. Because of this relation, the photon dispersion
relation can be expressed in terms of the single parameter $\xi_n^+$ which we denote as $\xi_n$
in the following.

Effective field theory predicts the relation $\eta^{\rm p,\,\pm}_{n}=(-1)^n \eta^{\rm e,\,\mp}_{n} $
between the LIV parameters for fermions and anti-fermions
for $n=1$ and $n=2$~\cite{Jacobson:2003bn, Mattingly:2008pw}.
It is, therefore, not necessary to introduce new parameters in the modified dispersion relation 
for positrons which can thus be written as
\begin{equation}
E_{\rm p,\,\pm}^2=p_{\rm p}^2+m_{\rm e}^2+(-1)^n \eta^{\rm e,\,\mp}_{n} p_{\rm p}^2\left(\frac{p_{\rm p}}{M_{\rm pl}}\right)^n\,.
\end{equation}
Thus, for the remainder of this paper we can restrict ourselves to the parameters
$\eta^{\rm e,\,\pm}_{n}$, for which we simply write $\eta^\pm_{n}$.

As a result, LIV modifications in the QED sector are described by three parameters
at a given suppression by the power $n=1,2$ of the Planck scale, namely by
one for photons ($\xi_n$) and two for electron and positron ($\eta_n^+$ and $\eta_n^-$).
Note, however, that in some particular cases, kinematics is governed by just one parameter for the pair: 
If the final state of a certain process consists of an electron-positron pair with
opposite helicity, due to the relation $\eta^{\rm p,\,\pm}_{n}=(-1)^n \eta^{\rm e,\,\mp}_{n} $,
only either $\eta^+_{n}$ or $\eta^-_{n}$ appears.

\section{Threshold Equations}
If Lorentz invariance is preserved, the main process influencing the propagation 
of UHE photons is pair production: Photons with energy higher than $m_{\rm e}^2/k_{\rm b}$
produce an electron-positron pair interacting with low-energy background photons of energy $k_{\rm b}$.
For interaction with CMB photons ($k_{\rm b}\simeq 6\times 10^{-4}$ eV) 
the lower threshold is $\sim4\times 10^{14}$ eV,
and the UHE photon flux is highly suppressed due to this interaction~\cite{Galaverni:2007tq}.

If the dispersion relations for photons and fermions are modified by LIV terms, 
then the lower threshold for this process can be modified and pair production 
can also become forbidden above a certain upper threshold.
Moreover, other processes, usually forbidden if Lorentz invariance is preserved, can become allowed.
In particular, photon decay ($\gamma\rightarrow e^-\,e^+$) and photon splitting ($\gamma\rightarrow N\,\gamma$) 
can play an important role in the propagation of UHE photons. 
Note that if pair production is forbidden above a certain upper threshold, than photon decay 
must also be forbidden: If the production of an electron positron pair is kinematically forbidden 
for two photons (pair production), then it will be forbidden also for a single photon (photon decay),
otherwise pair production on an infinitely soft background photon would be allowed~\cite{Maccione:2008iw}.

The characteristic timescale of these processes is relevant for their relative importance:
In particular, for UHE photons the photon splitting timescale is usually larger than the
propagation timescale~\cite{Maccione:2008iw,Jacobson:2002hd}, 
therefore, in the following we will focus mainly on pair production and photon decay processes.

Which and how many LIV parameters are involved in pair production or in photon decay processes
depends on the polarization of incoming photon(s) and 
on the helicity of the outgoing electron and positron, see Tab.~\ref{tab:Comb}.
Although right at the threshold, where the electron positron pair is produced without angular momentum,
only the $s$-wave contributes to the process and certain helicity combinations are forbidden, 
above the threshold also higher partial waves contribute and all possible channels (all partial waves) 
must be considered.

\begin{table*}[htbp]
\centering
 \begin{tabular}{|l||c|c|c||c|c|c|c|}
  \hline
     & UHE $\gamma$ &  $e^-$ & 	$e^+$  & $\left(\xi,\eta_{\rm e},\eta_{\rm p} \right)$ 
    &number of LIV param. & $s$-wave allowed for PP & $s$-wave allowed for PD\\
  \hline
   1 &    $+$       &    $+$   &     $+$   &  $\left(\xi_n ,\eta_n^+, (-1)^n\eta_n^- \right)$ 
   & 3 & No & Yes \\
  \hline
   2 &    $+$       &    $+$   &     $-$   &  $\left(\xi_n ,\eta_n^+, (-1)^n\eta_n^+ \right)$ 
   & 2 & Yes & No \\
  \hline
   3 &    $+$       &    $-$   &     $+$   &  $\left(\xi_n ,\eta_n^-, (-1)^n\eta_n^- \right)$  
   & 2 & Yes & No \\
  \hline
   4 &    $+$       &    $-$   &     $-$   &  $\left(\xi_n ,\eta_n^-, (-1)^n\eta_n^+ \right)$  
   & 3 & No & Yes \\
  \hline
   5 &    $-$       &    $+$   &     $+$   &  $\left((-1)^n \xi_n ,\eta_n^+, (-1)^n\eta_n^- \right)$  
   & 3 & No & Yes \\
  \hline
   6 &    $-$       &    $+$   &     $-$   &  $\left((-1)^n \xi_n ,\eta_n^+, (-1)^n\eta_n^+ \right)$  
   & 2 & Yes & No \\
  \hline
   7 &    $-$       &    $-$   &     $+$   &  $\left((-1)^n \xi_n ,\eta_n^-, (-1)^n\eta_n^- \right)$  
   & 2 & Yes & No \\
  \hline
   8 &    $-$       &    $-$   &     $-$   &  $\left((-1)^n \xi_n ,\eta_n^-, (-1)^n\eta_n^+ \right)$  
   & 3 & No & Yes \\
  \hline
 \end{tabular}
	\caption{Photon, electron and positron LIV parameters ${\rm \left(\xi,\eta_{\rm e},\eta_{\rm p} \right)}$ 
	appearing in the combination $\xi- \eta_{\rm p} y^{n+1}-\eta_{\rm e} \left(1-y\right)^{n+1}$ in the kinematic   
	equations~(\ref{eq:PP1}) and~(\ref{eq:PD1}) for pair production and photon decay, 
	respectively,  
	for	both polarizations of the UHE photon and for all helicity configurations of the electron-positron pair.
	Recall that for electrons, the sign index $\pm$ in $\eta_n^\pm$ refers to the helicity, whereas
  for positrons it refers to the inverse helicity. 
	For each combination we show the number of LIV parameters contributing to the kinematics, and whether
	conservation of total angular momentum allows the $s$-wave channel for pair production (PP),
	or for photon decay (PD) at threshold.}
	\label{tab:Comb}
\end{table*}

\subsection{Pair Production ($\gamma\,\gamma_{\rm b}\rightarrow e^-\,e^+$)}
Exact energy momentum conservation implies that
\begin{equation}
\left(\omega_\pm+\omega_{\rm b}\right)^2-\left({\bf k}+{\bf k_{\bf \rm b}}\right)^2=
\left(E_{\rm e,\,\pm}+E_{\rm p,\,\pm}\right)^2- \left({\bf p_{\bf \rm e}}+{\bf p_{\bf \rm p}}\right)^2\,.
\end{equation}
The left-hand side is maximized for a head-on collision of the two photons, 
while the right-hand side is minimized for parallel final momenta of the
pair~\cite{Jacobson:2002hd,Mattingly:2002ba}. 
Expanding in terms of the LIV parameters, and writing $p_{\rm e}=(1-y)k$ and $p_{\rm p}=yk$ 
as functions of the asymmetry $y$ in the final momenta, for right polarized photons
we obtain the equation~\cite{Galaverni:2007tq}:
\begin{eqnarray}
\label{eq:PP1}
& &\left[\xi_n-\left(-1\right)^n \eta_n^\mp y^{n+1}-\eta_n^\pm \left(1-y\right)^{n+1}\right]k^2\left(\frac{k}{M_{\rm pl}}\right)^n\nonumber\\
& & +4k k_{\rm b}-\frac{m_{\rm e}^2}{y(1-y)}=0\,,
\end{eqnarray}
where all four combinations of $\eta_n^\mp$ and $\eta_n^\pm$ can occur in the square
bracket. Note that due to the relation $\eta^{\rm p,\,\pm}_{n}=(-1)^n \eta^{\rm e,\,\mp}_{n} $,
for electrons, the sign index $\pm$ in $\eta_n^\pm$ refers to the helicity, whereas
for positrons it refers to the inverse helicity. As a consequence, equal sign indices in the two terms
correspond to opposite helicities for electron and positron, whereas opposite sign indices
correspond to equal helicities. For left polarized UHE photons $\xi_n$ 
must be replaced with $(-1)^n \xi_n$. The second term inside square brackets refers 
to positrons and the third one to electrons. By definition,
in the third term $\eta_n^+$ refers to electrons of positive helicity, and
$\eta_n^-$ to electrons of negative helicity. In contrast, in the second term
$\eta_n^+$ refers to positrons of negative helicity, whereas $\eta_n^-$
refers to positrons of positive helicity. Therefore,
Eq.~(\ref{eq:PP1}) reduces to Eq.~(3) of Ref.~\cite{Maccione:2008iw} in the
channel where electrons and positrons have opposite helicities, where kinematics
depends only on either $\eta_n^+$ or $\eta_n^-$. Alternatively, one also
obtains this equation in the channel where electrons and positrons have the
same helicities, if one assumes the relation $\eta_n^+=\eta_n^-$ between electron LIV
parameters with opposite helicity.

When all LIV parameters vanish, we find the usual Lorentz invariant lower
threshold ($k_{\rm LI}=m_{\rm e}^2/k_{\rm b}$) for a symmetric final configuration $y=1/2$, .

For given values for LIV parameters we determine numerically the
lower and upper thresholds of this process using Eq.~(\ref{eq:PP1}) and its derivative
with respect to $k$ and $y$~\cite{Jacobson:2002hd}.

Defining $x\equiv4 y(1-y) k/k_{\rm LI}$, Eq.~(\ref{eq:PP1}) can be rewritten as
\begin{equation}
\label{eq:PP2}
\alpha_n x^{n+2}+x-1=0\,,
\end{equation}
where, for right polarized photons,
\begin{equation}
\alpha_n \equiv\frac{\xi_n-\left(-1\right)^n \eta_n^\mp y^{n+1}-\eta_n^\pm \left(1-y\right)^{n+1}}
{2^{2\left(n+2\right)}y^{n+1}\left(1-y\right)^{n+1}}
\frac{m_{\rm e}^{2\left(n+1\right)}}{k_{\rm b}^{n+2} M_{\rm pl}^n} \,,
\end{equation}
and for left polarized photons $\xi_n$ must be replaced by $(-1)^n \xi_n$.

If $n\geq 1$, Eq.~(\ref{eq:PP2}) has at most two positive solutions, corresponding 
to a lower or an upper threshold. If there were more than two positive solutions, there would be 
two or more stationary points for $x>0$, but the derivative of Eq.~(\ref{eq:PP2}) vanishes
for
\begin{equation}
\left(n+2\right)\alpha_n x^{n+1}+1=0\,,
\end{equation}
and the solutions of this equation are
\begin{eqnarray}
x_s=\left[\left(n+2\right)\alpha_n\right]^{-\frac{1}{n+1}} \exp\left[i\frac{\pi+2 \pi s}{n+1}\right]\;\left(\mbox{if}\;\alpha_n>0\right)\\
x_s=\left[\left(n+2\right)\left|\alpha_n\right|\right]^{-\frac{1}{n+1}} \exp\left[i\frac{2 \pi s}{n+1}\right]\;\left(\mbox{if}\;\alpha_n<0\right)
\end{eqnarray}
where $s=0,\dots,n$.
These expressions are real and positive only for $s=0$, therefore
there cannot be more than one stationary point. This excludes the possibility that
there could be more than two thresholds for pair production. 

As in Ref.~\cite{Galaverni:2007tq}, we will argue that current upper limits on the
photon fraction of UHE cosmic rays in the energy range between $\simeq10^{19}\,$eV
and $\simeq10^{20}\,$ eV require that pair production has to be kinematically allowed
for both UHE photon polarizations shown in Tab.~\ref{tab:Comb},
otherwise at least one channel would be unabsorbed and one would expect $\gtrsim10$\%
photons. This will rule out certain ranges
in the parameter space of the three LIV parameters $\xi_n$ and $\eta_n^\pm$. However,
in order to be conservative, we will rule out only parameter combinations for which
the photon is stable. This is because for unstable photons, the absence of photons
in the observed ultra-high energy cosmic ray flux could be due to photon decay, $\gamma\rightarrow e^-\,e^+$,
at least as long as any electron-positron pairs in the decay products cannot
recreate a significant photon flux by inverse Compton scattering on the CMB. 
On the other hand, the observation of a UHE photon would rule out photon decay because
this process would occur on microscopic time scales once it is allowed.
We will, therefore, also consider photon decay in the following.

\subsection{Photon decay ($\gamma\rightarrow e^-\,e^+$)}
For the reaction $\gamma\rightarrow e^-\,e^+$ 4-momentum conservation implies:
\begin{equation}
\omega_\pm^2-k^2=
\left(E_{\rm e,\,\pm}+E_{\rm p,\,\pm}\right)^2- \left({\bf p_{\bf \rm e}}+{\bf p_{\bf \rm p}}\right)^2\,,
\end{equation}
and proceeding as for pair production we obtain for right polarized UHE photons:
\begin{eqnarray}
\label{eq:PD1}
& & \left[\xi_n-\left(-1\right)^n \eta_n^\mp y^{n+1}-\eta_n^\pm \left(1-y\right)^{n+1}\right]k^2\left(\frac{k}{M_{\rm pl}}\right)^n\nonumber\\
& & -\frac{m_{\rm e}^2}{y(1-y)}=0\,.
\end{eqnarray}
The corresponding equation for left polarized photons is obtained
by substituting $\xi_n$ with $(-1)^n \xi_n$.

Note that Eq.~(\ref{eq:PP1}) for pair production reduces to Eq.~(\ref{eq:PD1})
for photon decay when the energy of the background photon $k_{\rm b}$ vanishes.

Photon decay is kinematically forbidden in the Lorentz invariant case,
but it can become allowed, above a certain energy threshold, for certain values
of the LIV parameters. We again search numerically for this threshold by employing
Eq.~(\ref{eq:PD1}) and its derivatives with respect to $k$ and $y$.  

If a photon of a certain energy is detected, at least one photon polarization
must be stable, i.e. cannot decay into any helicity configuration of the final pair.

Eq.~(\ref{eq:PD1}) can be rewritten as:
\begin{equation}
\alpha_n x^{n+2}-1=0\,,
\end{equation}
and its solutions are of the form
\begin{eqnarray}
x_s&=&\alpha_n^{-\frac{1}{n+2}} \exp\left[i\frac{2 \pi s}{n+2}\right]\quad\left(\mbox{if}\;\alpha_n>0\right)\\
x_s&=&\left|\alpha_n\right|^{-\frac{1}{n+2}} \exp\left[i\frac{\pi+2 \pi s}{n+2}\right]\quad\left(\mbox{if}\;\alpha_n<0\right)
\end{eqnarray}
where $s=0,\dots,n+1$.
Note that these expressions give at most one positive solution $x_s$, 
therefore, there cannot be more than one threshold for photon decay.

\section{Constraints on Lorentz invariance violating terms}
Current upper limits on the photon fraction in the energy range between $\simeq10^{19}\,$eV and 
$\simeq10^{20}\,$eV already establish strong constraints on the LIV parameters 
in the cases $n=1$ and $n=2$~\cite{Galaverni:2007tq,Maccione:2008iw}.

If photon decay is forbidden, pair production must be
kinematically allowed for both high energy photon polarizations, 
otherwise the predicted photon flux would be too high.
According to the current non-detection of UHE photons~\cite{Risse:2007sd}
this leads to the condition that the lower threshold for pair production must
be below $\simeq10^{19}\,$eV and the upper threshold for pair production must
be above $\simeq10^{20}\,$eV~\cite{Galaverni:2007tq}.
At the threshold the pair can be produced only in $s$-waves, whereas higher partial waves are
forbidden. Above the lower threshold and below the upper threshold the pairs can also be produced
in higher partial waves which, therefore, also have to be considered. In fact, at energies
that are factors of a few away from the thresholds, say at $\simeq3\times10^{19}\,$eV, the pair
is produced with relative velocities not much smaller than the speed of light and, without
doing a detailed calculation, higher partial waves are thus not expected to be significantly suppressed.
Therefore, according to Tab.~\ref{tab:Comb}, all three LIV parameters enter the problem.
The experimental upper limits on the UHE photon flux require that a given combination
$(\xi_n,\eta_n^+,\eta_n^-)$ of LIV parameters is ruled out if at least one photon polarization
state is stable against decay and does not pair produce for any helicity configurations of the
final pair. Taking into account higher partial waves for pair production then leads to
conservative constraints because only these parameter combinations are ruled out that
do not lead to pair production into any of the final state configurations shown in
Tab.~\ref{tab:Comb}.

If a UHE photon were detected and its polarization were not measured, then
there should be at least one polarization state that is stable over
macroscopic time scales. Then the LIV parameter region where photon decay is
kinematically allowed for at least one helicity configuration of the final state
electron-positron pair, for both photon polarizations, would be ruled out.

Note that at the threshold for pair production, where only the $s$-wave channel is allowed,
according to Tab.~\ref{tab:Comb}, only opposite helicities for electron and positron
contribute and the kinematics depends on only one fermionic LIV parameter, either
$\eta_n^+$ or $\eta_n^-$. In contrast, at the threshold for photon decay, where the
photon, electron and positron momenta are all parallel, only equal helicities for electron and positron
contribute and the kinematics depends on all two fermionic LIV parameters,
since total angular momentum cannot be conserved if electron and positron have opposite helicity. 
For photon decay, therefore, even at the threshold only the assumption of an additional
relation between the electron LIV parameters for different helicities, e.g. $\eta^+_n=\eta_n^-$,
leads to a reduction of the kinematics to only one fermionic LIV parameter.

\subsection {Case $n=1$ -- $\mathcal{O}(E/M_{\rm pl})$ modifications of the dispersion relations}

\begin{figure}[htdb]
\includegraphics[width=8.6cm]{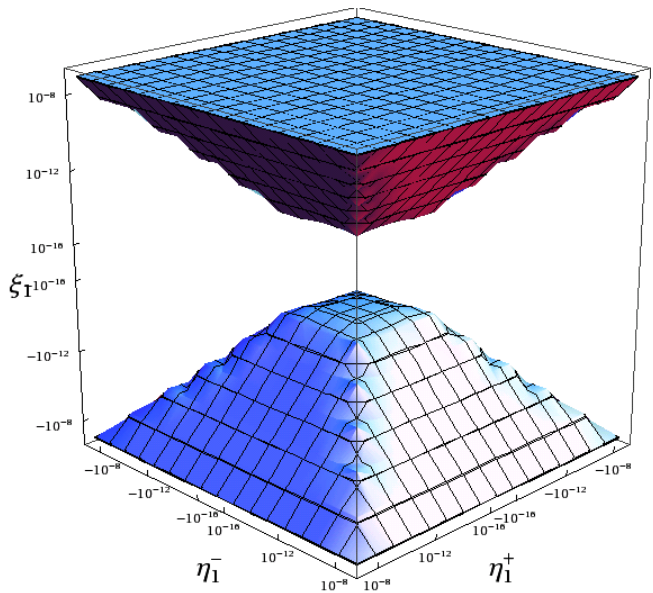}
\caption{Case $n=1$. Region excluded by present upper limits on the UHE photon flux 
($10^{19}\,\mbox{eV}\lesssim\omega\lesssim10^{20}\,\mbox{eV}$).
} 
\label{1PP_3d}
\end{figure}
\begin{figure}[htdb]
\includegraphics[width=8.6cm]{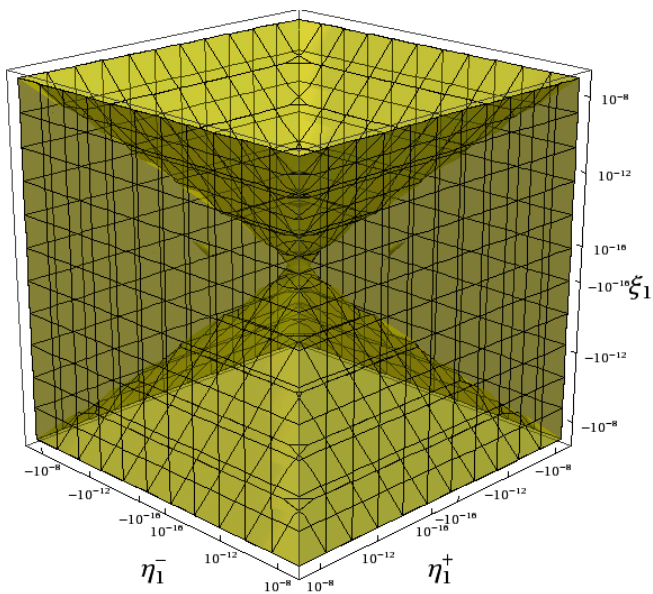}
\caption{Case $n=1$. Region excluded if a $10^{19}\,$eV photon were detected (shaded). The allowed
region has the shape of a double-cone centered at the origin and opening towards the
positive and negative $\xi_1$-directions.}
\label{1PD_3d}
\end{figure}
\begin{figure}[htdb]
\includegraphics[width=8.6cm]{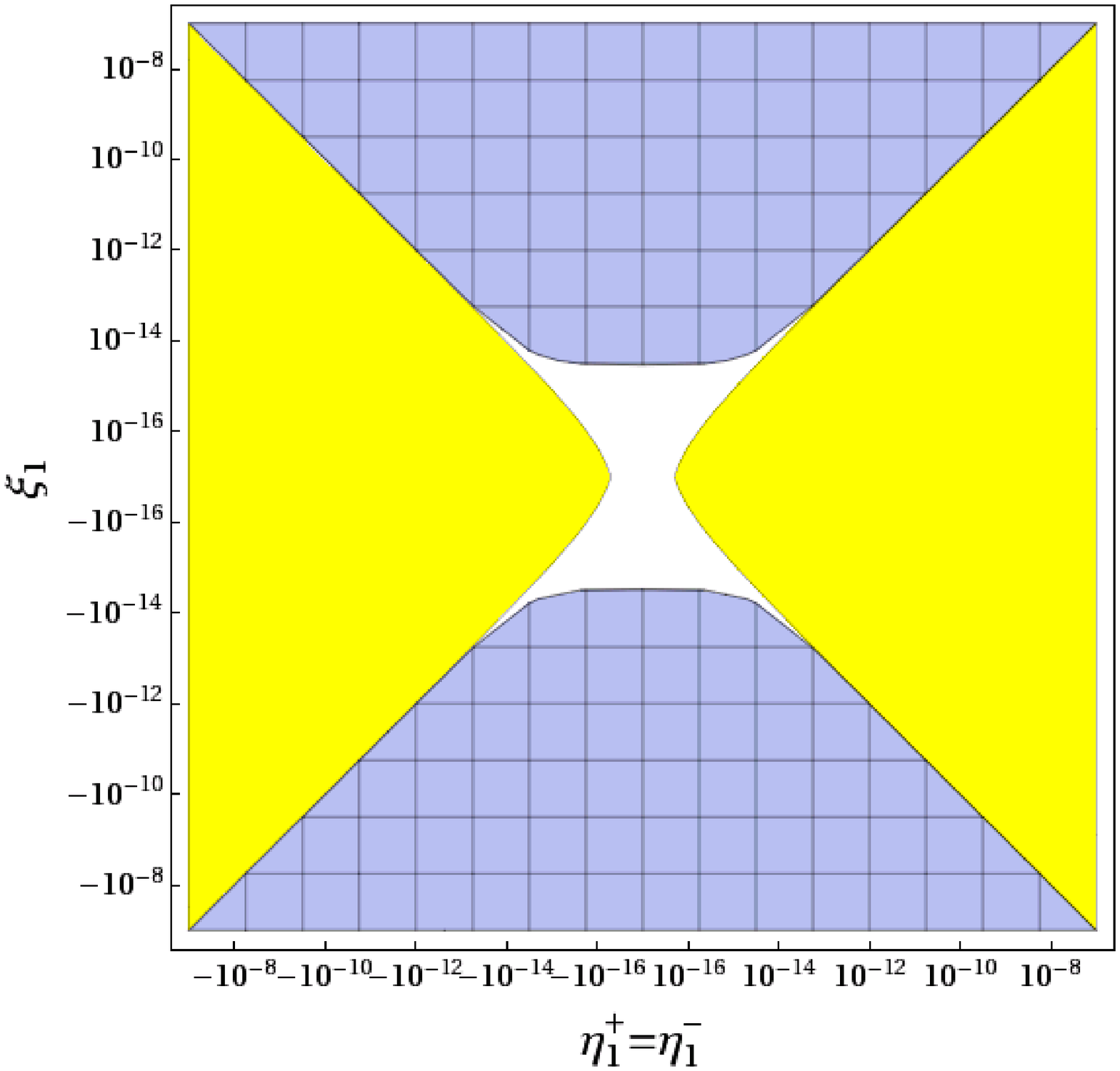}
\caption{Case $n=1$, $\eta_1^+=\eta_1^-$.  
Combined constraint using both the current upper limits on the photon fraction 
in the energy range between $10^{19}$ eV and  $10^{20}$ eV  (blue shaded, checkered region), 
and assuming that a $10^{19}$ eV photon were detected (yellow shaded region).} 
\label{1PPPD_2dA}
\end{figure}
\begin{figure}
\includegraphics[width=8.6cm]{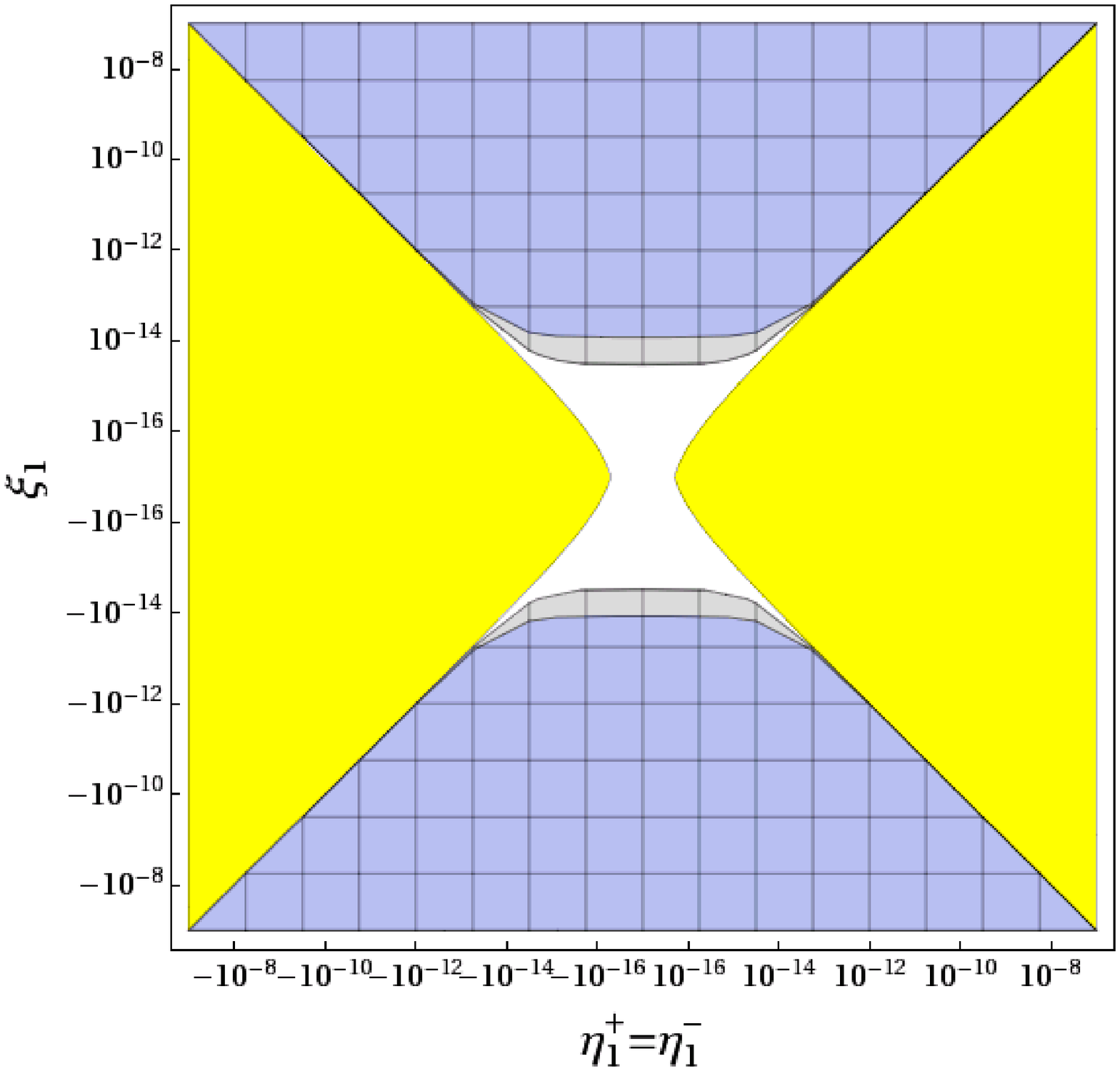}
\caption{Case $n=1$, $\eta_1^+=\eta_1^-$.  
Combined constraint using the current upper limits on the photon fraction 
in the energy range between $10^{19}$ eV and  $10^{20}$ eV  (gray plus blue shaded,
checkered regions),
in the energy range between $10^{19}$ eV and  $5\times10^{19}$ eV  (blue region),
and assuming that a $10^{19}$ eV photon were detected (yellow shaded region).} 
\label{1PPPD_2dA2}
\end{figure}

For first order suppression in the Planck scale,
the excluded LIV parameters resulting from the current non-detection of a photon component of
cosmic rays in the energy range between $10^{19}$ eV and $10^{20}$ eV are shown in Fig.~\ref{1PP_3d}.
The excluded parameter region is symmetric with respect to a sign change of the photon LIV parameter
$\xi_1$ because pair production must be allowed for both photon polarizations which correspond
to opposite signs of $\xi_1$ for $n=1$.
Note that if the absolute values of the LIV parameters $\eta_1^+$ and $\eta_1^-$ for electrons and
positrons are smaller than the one for the LIV parameter for photons, parameters of size
$\left|\xi_1\right|\gtrsim 10^{-14}$ are
ruled out in agreement with Ref.~\cite{Galaverni:2007tq}.

For the parameter range shown in Fig.~\ref{1PD_3d} photons of energy $E\sim10^{19}\,$eV
and of both polarizations would be unstable. Thus, if a $\sim10^{19}$ eV photon were detected
without determining its polarization, this parameter range would be excluded. The allowed
parameter range has the structure of a double-cone and is symmetric with respect to a sign
change of the photon LIV parameter $\xi_1$ because opposite photon polarizations correspond
to opposite signs of $\xi_1$ and within our conservative treatment only one photon polarization
needs to be stable. As a result, the region that would be excluded by a $\sim10^{19}$ eV photon
detection is also symmetric with respect to a sign change of $\xi_1$. The resulting constraints
would be very strong: If, for example, $\left|\xi_1\right|\ll\left|\eta_1^\pm\right|$, then
$\left|\eta_1^\pm\right|\gtrsim 10^{-16}$ would be excluded.

The sign and parameter combinations entering the kinematics in Tab.~\ref{tab:Comb} lead to additional
symmetries of parameter ranges excluded and allowed by pair production and/or photon
decay under sign changes of $\eta_1^+$ or $\eta_1^-$ and under exchange of $\eta_1^+$ and $\eta_1^-$.

As Figs.~\ref{1PP_3d} and~\ref{1PD_3d} show, combining both constraints from UHE photon
flux limits and detection of an UHE photon it would be possible to rule out all LIV parameters
of absolute value larger than $10^{-14}$.

Fig.~\ref{1PPPD_2dA} shows, for the same range of LIV parameters studied in Figs.~\ref{1PP_3d} and \ref{1PD_3d},
a typical two dimensional section, $\eta_1^+=\eta_1^-$, of the excluded regions. This section
is relevant if only one leptonic LIV parameter enters the kinematics Eqs.~(\ref{eq:PP1})
and~(\ref{eq:PD1}). According to Tab.~\ref{tab:Comb}
this occurs very close to the threshold where only $s$-waves contribute to pair
production and electrons and positrons thus have opposite helicity~\cite{Maccione:2008iw}.
Alternatively, the section $\eta_1^+=\eta_1^-$ is relevant also away from the threshold
if the general (restrictive) assumption is made that the LIV parameter
for a positive helicity electron is equal to the LIV parameter for a negative helicity electron
($\eta_1^+=\eta_1^-$).
Note, however, at least in the $n=1$ case considered, that the order of magnitude of the largest
LIV parameters allowed does not depend on any particular relation assumed between $\eta^+_1$ and $\eta^-_1$: 
Parameters of absolute value larger than $10^{-14}$ are always ruled out.

Fig.~\ref{1PPPD_2dA2} shows, for the case $\eta_1^+=\eta_1^-$, how the uncertainties in the
photon fraction limits influence 
the constraints on LIV parameters. Lowering the maximum energy, up to which we
consider the bounds on the photon flux meaningful, from $10^{20}$ eV to $5\times10^{19}$ eV, 
the excluded region is slightly reduced. 

\subsection {Case $n=2$ -- $\mathcal{O}(E^2/M_{\rm pl}^2)$ modifications of the dispersion relations}

\begin{figure}[htdb]
\includegraphics[width=8.6cm]{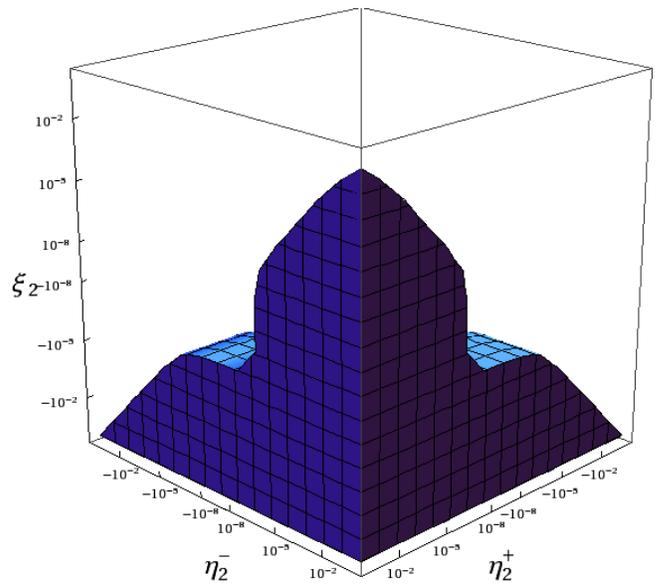}
\caption{Case $n=2$. Region excluded by present upper limits on the UHE photon flux 
($10^{19}\,\mbox{eV}\lesssim\omega\lesssim10^{20}\,\mbox{eV}$).
} 
\label{2PP_3d}
\end{figure}
\begin{figure}[htdb]
\includegraphics[width=8.6cm]{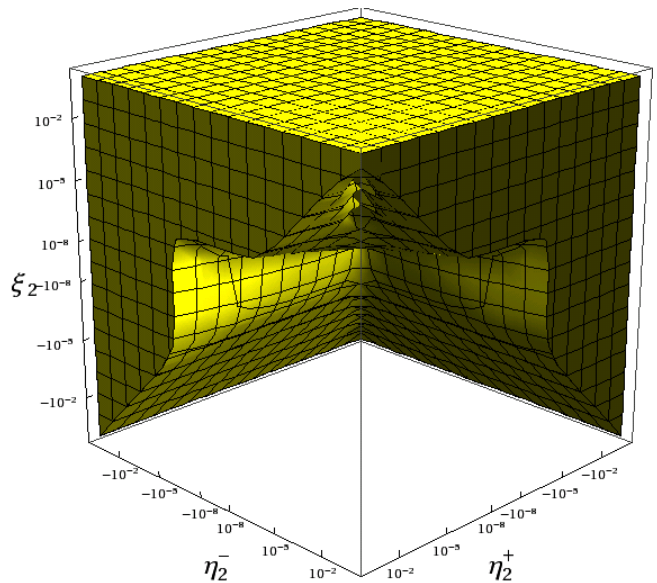}
\caption{Case $n=2$. Region excluded if a $10^{19}$ eV photon were detected.} 
\label{2PD_3d}
\end{figure}
\begin{figure}[htdb]
\includegraphics[width=8.6cm]{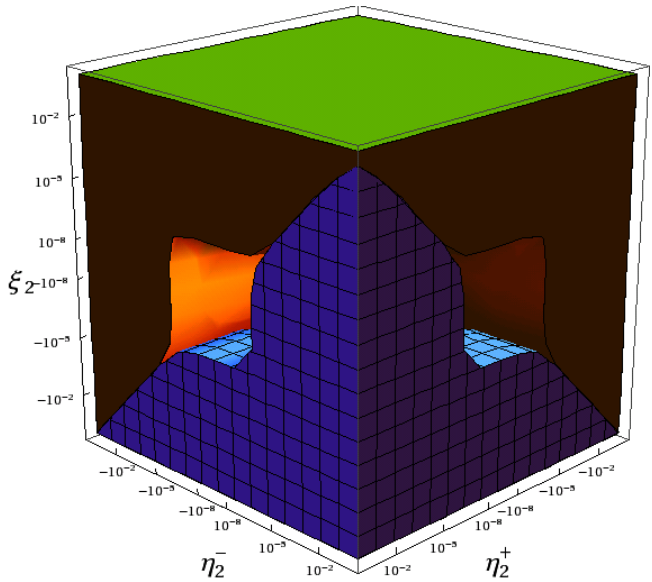}
\caption{Case $n=2$. Combined constraint using both the current upper limits on the photon fraction 
in the energy range between $10^{19}$ eV and  $10^{20}$ eV  (blue shaded, checkered region), 
and assuming that a $10^{19}$ eV photon were detected (uncheckered region).} 
\label{2PPPD_3d}
\end{figure}
\begin{figure}
\includegraphics[width=8.6cm]{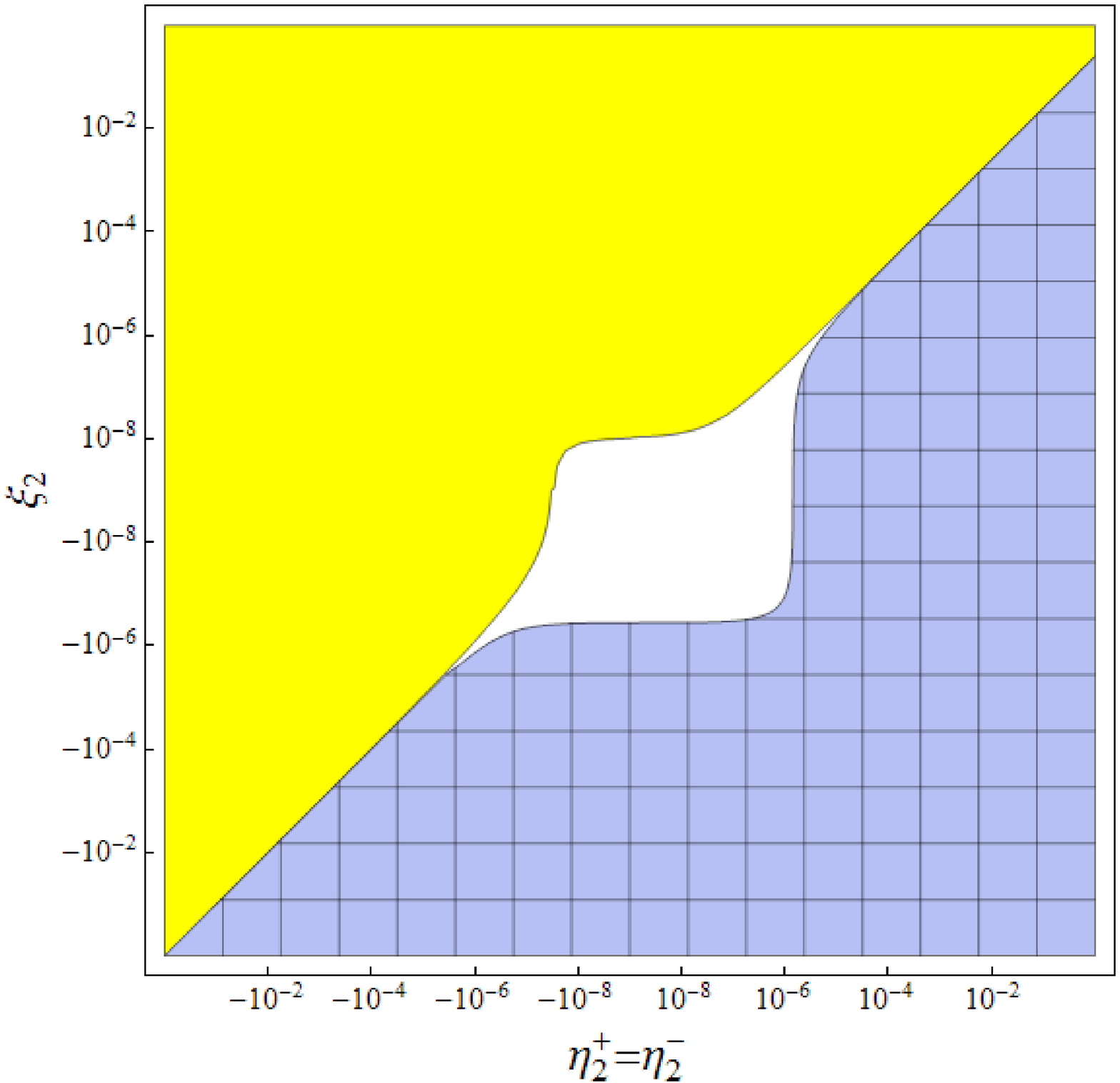}
\caption{Case $n=2$, $\eta_2^+=\eta_2^-$.  
Combined constraint using both the current upper limits on the photon fraction 
in the energy range between $10^{19}$ eV and  $10^{20}$ eV  (blue shaded, checkered region), 
and assuming that a $10^{19}$ eV photon were detected (yellow shaded region).} 
\label{2PPPD_2dB}
\end{figure}
\begin{figure}
\includegraphics[width=8.6cm]{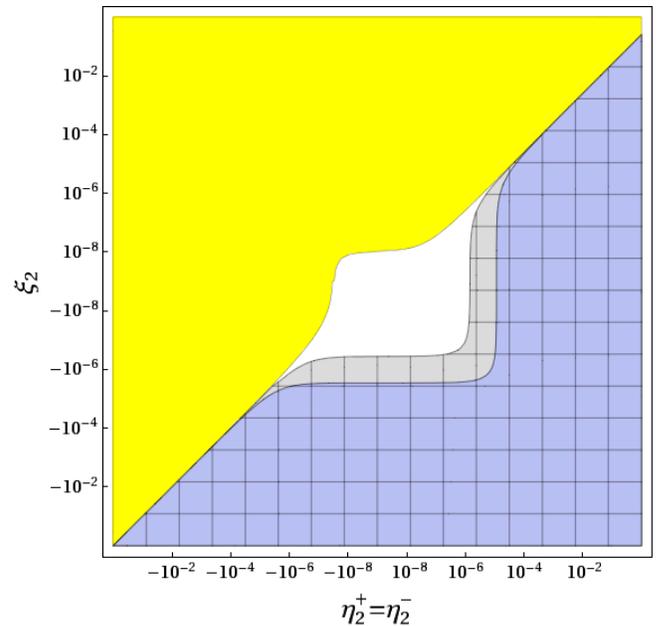}
\caption{Case $n=2$, $\eta_2^+=\eta_2^-$.  
Combined constraint using the current upper limits on the photon fraction 
in the energy range between $10^{19}$ eV and  $10^{20}$ eV  (gray plus blue regions),
in the energy range between $10^{19}$ eV and  $5\times10^{19}$ eV  (blue shaded, checkered region),
and assuming that a $10^{19}$ eV photon were detected (yellow shaded region).} 
\label{2PPPD_2dBb}
\end{figure}
\begin{figure}
\includegraphics[width=8.6cm]{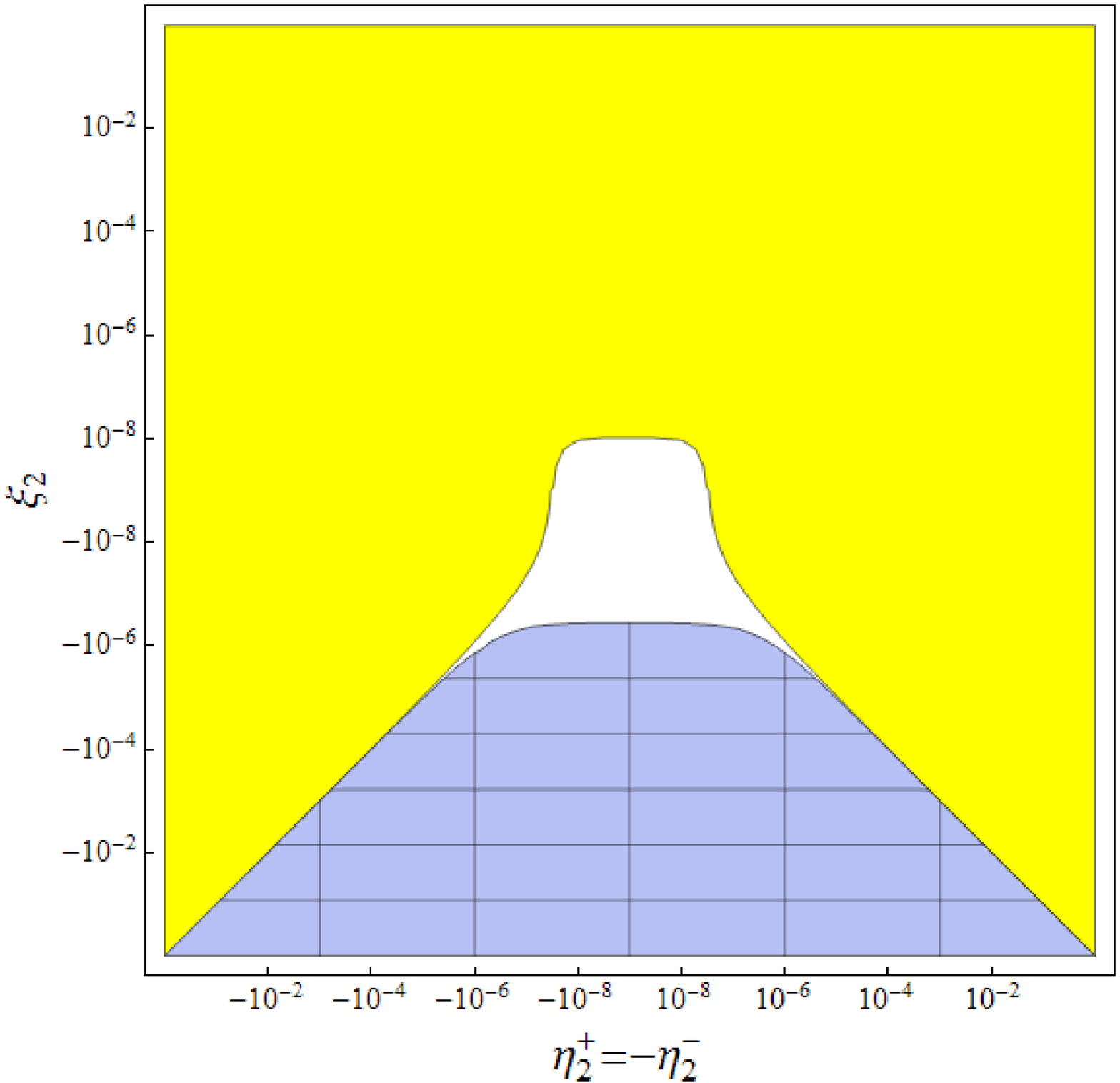}
\caption{Case $n=2$, $\eta_2^+=-\eta_2^-$.  
Combined constraint using both the current upper limits on the photon fraction 
in the energy range between $10^{19}$ eV and  $10^{20}$ eV  (blue shaded, checkered region), 
and assuming that a $10^{19}$ eV photon were detected (yellow shaded region).} 
\label{2PPPD_2dA}
\end{figure}
\begin{figure}
\includegraphics[width=8.6cm]{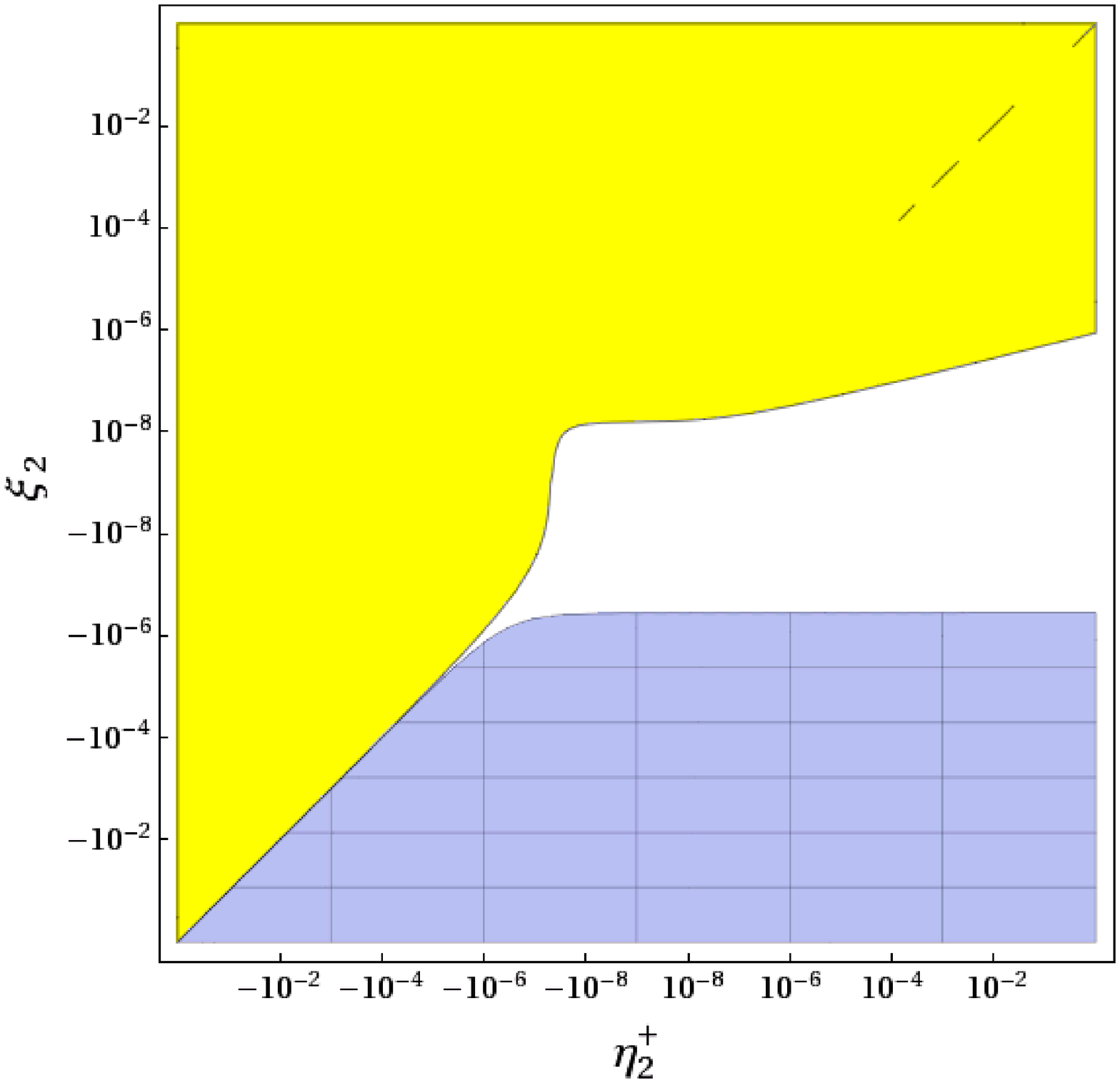}
\caption{Case $n=2$, $\eta_2^-=0$.  
Combined constraint using both the current upper limits on the photon fraction 
in the energy range between $10^{19}$ eV and  $10^{20}$ eV  (blue shaded, checkered region), 
and assuming that a $10^{19}$ eV photon were detected (yellow shaded region).} 
\label{2PPPD_2dC}
\end{figure}

The relations between the LIV parameters of photons with opposite polarization ($\xi_2^+=\xi_2^-$), and 
between the LIV parameters of electrons and positrons ($\eta^{\rm p,\,\pm}_{2}=\eta^{\rm e,\,\mp}_{2} $) have opposite sign with respect to the $n=1$ case, because
second order Planck scale suppressed modifications of the dispersion relations correspond to CPT-even
LIV operators. Therefore, for given polarizations of the incoming photon and given
helicities of the final electron positron pair, the signs of the parameters appearing in the kinematic equations 
for pair production and for photon decay are different from the case $n=1$, see Tab.~\ref{tab:Comb}.   
As a result, the region of the LIV parameter space ruled out for $n=2$ is not only smaller
than the one for $n=1$, but also has a different shape. In particular, it does not exhibit
the symmetry under sign changes
of either one of the three LIV parameters $\xi_2$, $\eta_2^+$ or $\eta_2^-$.

The current non-detection of UHE photons require that photons in the energy range between
$10^{19}$ eV and $10^{20}$ eV are subject to pair production, and the resulting excluded
range of $n=2$ LIV parameters is shown in Fig.~\ref{2PP_3d}. Note that if $\left|\eta_2^+\right|$
and $\left|\eta_2^-\right|$ for electrons and positrons are smaller than $\left|\xi_2\right|$,
then $\xi_2\lesssim-10^{-6}$ is ruled out in agreement with Ref.~\cite{Galaverni:2007tq}.

Fig.~\ref{2PD_3d} shows the excluded region if a $10^{19}$ eV photon were detected in the
future, and Fig.~\ref{2PPPD_3d} represents the combination of the two regions that would then
be excluded.

Differently from the $n=1$ case the excluded region does not surround the origin in all directions, 
therefore these two conditions do not rule out all LIV parameters larger than a certain value.
The shape of the excluded region now strongly depends on the particular relation between the 
electron LIV parameters. We consider here, for the same range of LIV parameters, 
three different two dimensional sections of the excluded region of Fig.~\ref{2PPPD_3d},
namely $\eta_2^+=\eta_2^-$, $\eta_2^+=-\eta_2^-$, and $\eta_2^-=0$.

The first one, shown in Fig.~\ref{2PPPD_2dB}, corresponds to the particular case 
where the LIV parameters for electrons do not depend on helicity ($\eta_2^+=\eta_2^-$),
such that only one leptonic LIV parameter enters the kinematics Eqs.~(\ref{eq:PP1})
and~(\ref{eq:PD1}). As for the case $n=1$, Tab.~\ref{tab:Comb} shows that
this section is relevant very close to the threshold where electron and positron have opposite
helicity since only $s$-waves contribute to pair production~\cite{Maccione:2008iw}.
In this case all LIV parameters of absolute value larger than $\sim 10^{-6}$ are ruled out 
in agreement with Ref.~\cite{Maccione:2008iw}. Away from the threshold, the section
$\eta_2^+=\eta_2^-$ is relevant only under the restrictive assumption that the electron
LIV parameters are equal for both polarizations.
For this particular section we also estimate in Fig.~\ref{2PPPD_2dBb} how uncertainties on the energy range of
the upper limits on the flux of UHE photons modify the constrained region.

An orthogonal cut corresponds to the case where the LIV parameter for positive helicity electrons 
is opposite equal to the LIV parameter for negative helicity electrons, $\eta_2^+=-\eta_2^-$,
see Fig.~\ref{2PPPD_2dA}. The shapes of the two excluded regions are modified with respect
to Fig.~\ref{2PPPD_2dB}:
The region of the LIV parameter space ruled out by the detection of a $10^{19}$ photon increases, 
while the region excluded by the current upper limits on the flux of UHE photon decreases.
However, also in this case, all LIV parameters of absolute value larger than 
$\sim 10^{-6}$ can be ruled out. 

Fig.~\ref{2PPPD_2dC} represents the excluded region in the $\eta_2^-=0$ plane:  
The shapes of the two excluded regions change, moreover, it is no more possible 
to rule out all LIV parameters of absolute value larger than $\sim 10^{-6}$.
If $\left|\xi_2\right|\lesssim10^{-6}$, arbitrarily large $\left|\eta_2^+\right|$ are
currently not excluded, and even if a $\sim10^{19}\,$eV photon were detected in the
future, arbitrarily large positive $\eta_2^+$ could still not be excluded.

Note that this is true not only in the $\eta_2^-=0$ plane, but whenever the
absolute values of the LIV parameters of both the
photon ($\xi_2$) and of one of the lepton sector ($\eta_2^+$ or $\eta_2^-$) 
are smaller than $\sim 10^{-6}$, see Fig.~\ref{2PPPD_3d}: It is then no more possible 
to exclude all LIV parameters with modulus larger than a certain threshold.

As an application, the current constraints based on the nondetection of UHE photons rule out any possible interpretation of flares of the active galaxy Markarian 501 in terms of quantum gravity
effects~\cite{Albert:2007qk} within the effective field theory approach with exact energy-momentum conservation assumed in the present work. Such an interpretation would be based on
an energy dependent index of refraction in vacuum such that the speed of light is
modified to $v(\omega)=1+\xi_1(\omega/M_{\rm Pl})+\xi_2(\omega/M_{\rm Pl})^2$ and would
require $\xi_1\lesssim-25$ or $\xi_2\lesssim-3\times10^{16}$, clearly ruled out by
our constraints. Note, however, that the constraints obtained here do not apply to particular scenarios, such as quantum-gravitational foam models, in which energy fluctuates due to non-trivial particle recoil
off excitations in the string/D-particle foam~cite{Ellis:2000sf,Ellis:2008gg}.

\section{Conclusions}

In the present paper we derived general constraints on LIV dispersion relations in the QED sector 
from the propagation of UHE photons. 
The kinematics of pair production and photon decay is discussed in terms of all three LIV 
parameters $\xi_n$, $\eta_n^+$ and $\eta_n^-$ entering in the dispersion relations
of photons, electrons and positrons, both for linear ($n=1$) and for second order ($n=2$) 
suppression with the Planck scale.

The upper limits on the flux of UHE photons require that combinations of the LIV parameters 
for which the UHE photons are stable and cannot pair produce on low-energy (e.g. CMB) photons 
are excluded. Similarly, the detection of photons of $\sim10^{19}\,$eV would exclude those
combinations of the LIV parameters for which both photon polarizations are unstable.

For terms in the dispersion relation linearly suppressed by the Planck scale the resulting
constraints are very strong: Using the non-detection of an UHE photon flux and anticipating
the detection of a $\sim10^{19}\,$eV photon, it will be possible to exclude all LIV parameters
with absolute value $\gtrsim 10^{-14}$.

In contrast, in the $n=2$ case, the LIV parameter region excluded using these 
arguments based on UHE photon propagation does not completely surround the origin.
If UHE photons are eventually detected, the maximum absolute value allowed for LIV parameters
in the photon sector will be $\sim 10^{-6}$, whereas currently only values smaller than
$\sim -10^{-6}$ are ruled out. However, even if UHE photons were detected,
constraints on the electron parameters can be evaded for some particular combinations:
If, for example, the modulus of one of the two parameters $\eta_2^+$ or $\eta_2^-$ is smaller
than $\sim 10^{-6}$, then the modulus of the other parameter is constrained neither by the
upper limit on the UHE photon flux nor by a putative future detection of a $\sim10^{19}\,$eV photon.
However, the case where the moduli of both $\eta_2^+$ and $\eta_2^-$ are $\gtrsim 10^{-6}$ can
still be excluded once UHE photons are detected.

\section*{Acknowledgements} 
We are grateful to Stefano Liberati and Luca Maccione for enlightening discussions.
G.S. acknowledges partial support by the DFG (Germany) under grant SFB-676.
M.G. is partially supported by INFN IS PD51.

\end{document}